\documentstyle[aps,aipbook,floats,epsfig]{revtex}

\newcommand{\ho}{\hspace*{1.5cm}}
\input psfig
\begin{document}

\title{Static Quantities of the W bosons in the MSSM}

\author{A.B.Lahanas$^*$}\thanks{My thanks are due to the organizers
for inviting and giving me the opportunity to participate in the Symposium,
 and also to the UCLA Physics Department for the hospitality extended to me
during my stay in Los Angeles. Work supported in part by EEC contract
SCI-CT92-0792.} \address{University of ATHENS, Physics Department \\
Nuclear and Particle Physics Section \\
Athens 157 71 Greece}

\maketitle

\begin{abstract}
I discuss the static quantities of the W boson, magnetic dipole
and electric quadrupole moments, in the context of the minimal supersymmetric
standard model,  in which supersymmetry is broken by soft terms
$A_o,\, m_o,\, M_{1/2}$. Following a renormalization group analysis it is
found that the supesymmetric values of $\Delta k_{\gamma}$
and $\Delta Q_{\gamma}$ can be largely different, in some cases, from the
standard model predictions but of the same order of magnitude for values  of
\quad $A_0,m_0,M_{1/2} \leq {\cal O}(1 TeV) $. Therefore possible
supersymmetric structure can be probed provided the accuracy of measurements
for $\Delta k_{\gamma}$, $\Delta Q_{\gamma}$ reaches $10^{-2}-10^{-3}$
and hence hard to be detected at LEP2. In cases where
$M_{1/2} \ll  A_0,m_0 $, the charginos and neutralinos may give substantial
contributions saturating the LEP2 sensitivity limits. This occurs when their
masses $m_{\tilde C},m_{\tilde Z}$ turn out to be both light satisfying
$m_{\tilde C}+m_{\tilde Z} \simeq M_W $. However these extreme cases
are perturbatively untrustworthy and besides unnatural for they occupy a
small region in the parameter space.
\end{abstract}

\section*{Introduction}

{\bf Supersymmetry} is a reasonable extension of the SM , theoretically
motivated but without any direct experimental confirmation for its existence
as yet. The recently revived interest in supersymmetric theories derives
from the fact that high  precision measurements of the SM parameters
at LEP $e^{+}e^{-}$
CERN collider shows that SU(3),SU(2),U(1) gauge couplings merge at a single
point at energies  $\sim 10^{16} GeV$ if supersymmetry is adopted with
an effective SUSY breaking scale $M_S$ \cite{ekn}
\begin{eqnarray}
M_Z < M_S <1 TeV   \nonumber
\end{eqnarray}
In order to produce SUSY particles with such large masses at observable rates
high energies and luminocities are required and the question is if there are
signals for SUSY below the supersymmetric particle thresholds.

The  three gauge boson vertex will be probed in future experiments with
high accuracy  and it is perhaps a good place to look for supersymmetric
signatures.In particular
the static quantities of the W-boson are affected by the radiative effects
which are due to  supersymmetric particles and deviations from the Standard
Model predictions are expected.\\
{\bf Are these deviations detectable ? How they depend on the effective
SUSY scale ?}\\ To answer this
within the context of the $MSSM$  requires  a systematic analysis
in which alllimitations imposed by the RG  and the radiative symmetry
breaking are  duly taken into account.
\section*{The MSSM}

The MSSM is  the minimal  extension of the SM in that
it is based on the gauge group $SU(3) \times SU(2) \times U(1)$
and has the minimal physical content. It involves two Higgs multiplets
${\hat H}_{1,2}$, and the minimum number of chiral
quark and lepton multiplets to accomodate the matter fermions
$(\hat Q, {\hat U}^c ,{\hat D}^c ,\hat L ,{\hat E}^c  )$. Its Lagrangian  is
%
$$
{\cal L}={\cal L}_{{SUSY}}+{\cal L}_{{soft}}.  \nonumber
$$
%
$ {\cal L}_{{SUSY}}$ is its supersymmetric part derived from
a superpotential ${\cal W}$ bearing the form
%
$$
{\cal W}=h_U \hat{Q} \hat{H}_2 \hat{U}^c
       + h_D \hat{Q} \hat{H}_1 \hat{D}^c
        + h_E \hat{L} \hat{H}_1 \hat{E}^c
        + \mu \hat{H}_1 \hat{H}_2  \nonumber
$$
and  $ {\cal L}_{{soft}}$ is its supersymmetry breaking part given by
\begin{eqnarray}
 - {\cal L}_{{soft}} &=&{\sum_{i}} m_i^2 |\Phi_i|^2    \nonumber      \\
        & +& (h_U A_U Q H_2 U^c
         + h_D A_D Q H_1 D^c
         + h_E A_L L H_1 E^c+h.c.)     \nonumber      \\
                         &+&(  \mu B H_1 H_2 +h.c.)
         +\frac {1}{2} \sum_a M_a {\bar{\lambda}}_a \lambda_a .\nonumber
\end{eqnarray}
The sum extends over all scalar fields involved and all  family indices
have been suppressed \cite{nilles}.

All soft scalar masses $m_i$ , gaugino masses  $M_a$,
and trilinear scalar couplings $A_{U,D,L}$ are assumed equal at the
unification scale, that is we adopt universal boundary conditions as
suggested by grand unification and absence of FCNC.
$$
m_i=m_0 \, , \, A_{U,D,L}=A_0  \, , \,  M_a =M_{1/2}   ( at \, M_{GUT} )
$$
%
\noindent
This choice  parametrizes our ignorance concerning the origin of
the supersymmetry  breaking terms in the most economical way but it is in no
way mandatory. The sparticle mass spectrum is completely known once
all soft SUSY breaking and mixing parameters at the unification
scale $M_{GUT}$ are given as well as the top Yukawa coupling.
The number of parameters is reduced to five if we make use
of the fact that $M_z =91.18 GeV$. A convenient choice is to take as
independent parameters :
\begin{eqnarray}
m_0 \quad ,\quad M_{1/2}\quad  ,\quad A_0\quad,\quad m_t(M_z)\quad ,\quad
 { {\tan \beta}(M_z)=} {<H_2> \over <H_1>}     \nonumber
\end{eqnarray}
Then by running the RGE`s of all couplings and masses involved the full set
of parameters down at energies  $\sim M_z$ is known and predictions can be
made. There are some subtleties in this approach which are associated with
the breaking of the electroweak symmetry ,which takes place via radiative
corrections, the appearance of particle thresholds etc. which affect the low
energy predictions for the sparticle mass spectrum but these in no way affect
the static quantities of the W boson at the one loop order.

A typical mass spectrum is shown in table \ref{tab1} where the one loop
corrections to the Higgs particles, due to the heavy top and stop sector ,
have been taken into account. As is well known these yield large radiative
corrections especially to the lightest of the neutral Higgses involved.
\begin{table}
\caption{A typical mass spectrum of the MSSM for the inputs shown below}
\label{tab1}
\begin{tabular}{cc}
{$ m_t=170$} \hspace{3mm} ,\hspace{3mm} {$ \tan{\beta}=2.1$ } &
     {$ A_0,m_0,M_{1/2}$} \hspace{3mm}:\hspace{3mm} 500, 500, 75 \\
\tableline
   Particle      &  Physical mass \, (case $\mu > 0)$       \\
\tableline
Top : \,$M_t$           & 174.6              \\
\tableline
Higgses  & \\
$H_{\pm}$   & 786.4             \\
$H_{o}$        & 784.1              \\
$A$         & 782.3              \\
$h_o$         & 88.4       \\
\tableline
Squarks         &               \\
${\tilde u}_L ,{\tilde c}_L  $  &    524.3    \\
${\tilde u}_R ,{\tilde c}_R  $    &   523.0    \\
${\tilde d}_L ,{\tilde s}_L  $    &   527.8     \\
${\tilde d}_R ,{\tilde s}_R  $    &   523.9   \\
${\tilde t}_1 ,{\tilde t}_2  $  &   442.2 , 144.0   \\
${\tilde b}_1 ,{\tilde b}_2  $&   523.9 , 387.7   \\
\tableline
Sleptons \, : ${\tilde e}_L ,{\tilde \nu}_L ,{\tilde e}_R$   &
504.0 , 502.0 , 500.4  \\
\tableline
Gluinos \, :      ${\tilde g } $  &   189.0  \\
\tableline
Neutralinos \, : ${\tilde Z_{1,2,3,4} } $  &27.0 , 52.4 ,  503.7 , 489.1  \\
\tableline
Charginos \, : ${\tilde C_{1,2} }  $    &50.7 , 501.5
\end{tabular}
\end{table}
\section*{Static Quantities of the W -boson in the MSSM}
\subsection*{The $ WW\gamma$ vertex}
\begin{figure} 
\begin{center}
\mbox{\psfig{file=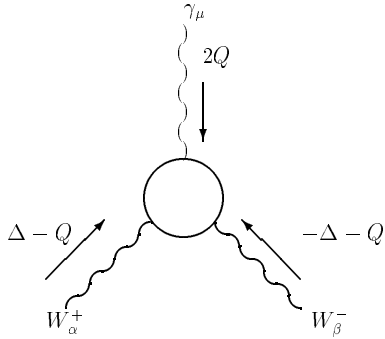,height=10cm}}
\end{center}
\caption{Kinematics of the $WWV$ vertex}
\label{fig1}
\end{figure}

The most general form of the
$WWV$ vertex ($V=\gamma, Z$), with the two W's on shell and neglecting the
scalar components of the boson $V$, is \cite{gounar}
\begin{eqnarray}
\Gamma_{\mu \alpha \beta}^{V}& =
&-ig_{V} \, \{ \,  {f_V}[2g_{\alpha \beta}{\Delta_\mu}+
       4(g_{\alpha \mu}{Q_\beta}-g_{\beta \mu}{Q_\alpha})]+ \nonumber\\
   &   &2 \, {\Delta k}_V \,
                          (g_{\alpha \mu}{Q_\beta}-g_{\beta \mu}{Q_\alpha})+
       4 \, {\frac  {{\Delta Q}_V}  {M_{ W}^{ 2}}} \, {\Delta_\mu}
       ({Q_\alpha}{Q_\beta}-{\frac {Q^{ 2}}{2}} g_{\alpha \beta})\}+...
\nonumber  \\
  &      & ( g_\gamma=e \, , \, g_Z=e\,cot{\theta_W} )   \nonumber
\end{eqnarray}
( ellipsis are $C$ and $CP$ odd terms ) \\
The labelling of the momenta and Lorentz indices is as shown in
figure \ref{fig1}.

${{\Delta k}_V}(Q^2)$,${{\Delta Q}_V}(Q^2)$ are functions of $Q^2$.
The static quantities of the $W$ boson magnetic dipole $\mu_W$ and electric
quadrupole $Q_W$ moments are related to these by
\footnote  {In other schemes in which parametrization in terms of $k_\gamma,
{\lambda}_\gamma$ is prefered:
$\mu_{ W}={\frac {e}{2M_{ W}}}\,(1+k_\gamma+{\lambda}_\gamma)$,
$Q_{ W}=-\,{\frac {e}{M_{ W}^2}}\,( k_\gamma-\lambda_\gamma  )$ } ,
%
%
\begin{eqnarray}
\mu_{ W}={\frac {e}{2M_{ W}}}\,(2+ {\Delta \kappa_{\gamma}}(0)) ,\quad
Q_{ W}=-\,{\frac {e}{M_{ W}^2}}\,
(1+{\Delta \kappa_{\gamma}}(0)+{\Delta Q_{\gamma}}(0)  )
\nonumber
\end{eqnarray}
 ${{\Delta k}_V}(Q^2),{{\Delta Q}_V}(Q^2)$ receive contributions from
radiative corrections due to the SM itself as well as from possible existence
of new physics which opens at some scale  $\Lambda>G_{F}^{-{\frac {1}{2}} }$.
In table \ref{tab2} we display the various sectors contributions to these
quantities in the $SM$ and $MSSM$.
\begin{table}
\caption{Particle  contributing to the static quantities of
the W boson}
\label{tab2}
\begin{tabular}{cc}
\ho   SM      & MSSM   \ho    \\
\tableline
\ho $ Gauge\quad bosons $      & $   Gauge\quad bosons$ \ho \\
\ho $ Matter\quad fermions $     & $ Matter\quad fermions  $ \ho       \\
\ho $ 1\, physical\quad Higgs   $ & $5\quad physical\quad Higgses  $ \ho \\
                   &$ {\tilde q} \, ,\, {\tilde l} $  \ho           \\
       &$ {\tilde Z }  \, , \,   {\tilde C}   $ \ho  \\
\end{tabular}
\end{table}
\subsection*{SM calculations}
Within the $SM$ ${{\Delta k}_\gamma}(0),{{\Delta Q}_\gamma}(0)$
were first calculated long time ago
by Bardeen,Gastmans and Lautrup, \cite{bard}.
The effect of the heavy fermion family (t,b) was subsequently  discussed
by Couture and Ng {\cite{cout}}. The form factors ${{\Delta k}_V}(Q^2)$,
${{\Delta Q}_V}(Q^2)$ have been also  calculated {\cite{argy}}
and their $Q^2$ dependence has been studied in detail. In that work it was
found that as  $Q^2$ grows  ${{\Delta k}_V}(Q^2)$ increases , violating
unitarity , and has  singular infrared (IF) behaviour  .
This reflects the fact that away from $Q^2=0$ the results are not gauge
independent. Actually the calculations in that reference were performed in
the 't Hooft-Feynman ($\xi=1$) gauge. In order to get gauge independent
results additional contributions stemming from box diagrams  have to be added
as was noted by Papavassiliou and Phillipides {\cite{papa}}.
%
\subsection*{SUSY calculations}
${{\Delta k}_\gamma}(0),{{\Delta Q}_\gamma}(0)$ \hspace{3mm}
have been also calculated in supersymmetric  versions of the SM.
Bilchak, Gastmans and Van Proyen  \cite{proy},
studied  ${{\Delta k}_\gamma}(0)$,${{\Delta Q}_\gamma}(0)$ in a particular
supersymmetric model in which electroweak symmetry is broken through a
singlet which gets nonvanishing v.e.v. SUSY however remains unbroken
in this model. Aliev \cite{ali},
dealt with the MSSM in which SUSY is broken by the appearance of soft terms
$A_0,B_0,m_0,M_{1/2}$ . However no renormalization group  analysis is
presented in that paper ; results are
only given in a particular case which is actually the supersymmetric limit
of the MSSM ,that is no soft SUSY breaking terms and absence of
Higgsino mixing parameter. It also seems that
the  contributions of the sensitive Neutralino-Chargino sector presented
in that reference are incorrectly given.
Couture,Ng,Hewett and Rizzo \cite{hewett},
did a more systematic analysis ; however the constraints imposed
by the Renormalization Group study of the MSSM , especially those from the
radiative breaking of the EW symmetry, have not been considered. Also mixings
of the  various sparticles occurring after electroweak breaking takes place
have been ignored.
In a more recent paper \cite{spa}, we
systematically analyzed the static quantities $\mu_W,Q_W$
or equivalently ${{\Delta k}_\gamma}(0)$,${{\Delta Q}_\gamma}(0)$  in the
context of the MSSM as functions of the soft SUSY breaking parameters
$A_0,B_0,m_0,M_{1/2}$ and the top quark mass . We followed a Renormalization
Group (RG) analysis and took into account all constraints imposed by the
radiative breaking scenario. The contributions of the various sectors
involved are as follows:
\subsubsection*{Gauge Bosons}
In units of $ {g^2}/{16{\pi^2}}\simeq 2.6 \times 10^{-3} $ and
for $Q^2=0$ the gauge boson contributions
to  ${{\Delta k}_\gamma}(0)$,${{\Delta Q}_\gamma}(0)$  are \cite{bard},
\begin{eqnarray*}
&&    \gamma\,:\quad   \Delta k_{\gamma}= {20 \over 3}{\sin^2}{\theta_W}
             \,\,,\,\, \Delta Q_{\gamma}={4 \over 9}{\sin^2}{\theta_W}   \\
&&   Z\,:\quad    \Delta k_{\gamma}={20 \over {3R}}-{5 \over 6}+
                 {1 \over 2}  \int_{0}^{1}dt
                 {t^4+10t^3-36t^2+32t-16  \over
                 t^2+R(1-t)  }   \\
&& \quad \,\, \quad  \Delta Q_{\gamma}=({8 \over {3R}}+{1 \over 3})
  \int_{0}^{1}dt     { t^3(1-t)  \over   t^2+R(1-t)  }   \\
&&  \quad \,\, \quad  (\,\,R\,=\,{(M_Z/M_W)}^2\,)  \\
\end{eqnarray*}
These result to
$$ {{\Delta k}_\gamma}(0) =1.18\quad,\quad {{\Delta Q}_\gamma}(0)=.235 $$
,in units of ${g^2}/{16{\pi^2}}$.
For nonvanishing $Q^2$ the Pinch Parts of the box graphs should be
included in order to get gauge independent results as already discussed.
\subsubsection*{Matter Fermions}
Matter fermions are the same in both SM and MSSM and such contributions have
been calculated. However we think there is a sign error in the original paper
of Bardeen et al which has been propagated in all following references
\cite{spa}. This has been also noted independently of us by  Culatti
\cite{cula}.

\begin{figure} 
\begin{center}
\mbox{\psfig{file=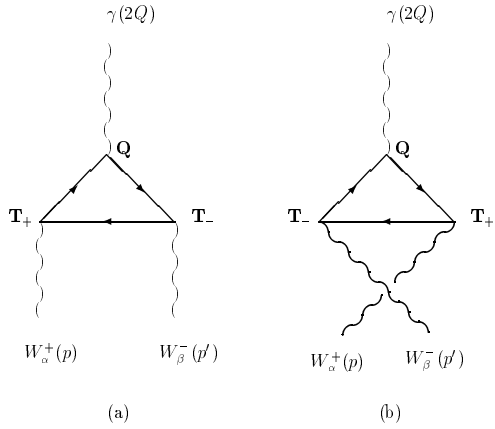,height=9cm}}
\end{center}
\caption{Triangle fermion graphs contributing to the magnetic dipole
and electric quadrupole moments.$Q,T_+,T_-$ denote electric  charge
and isospin raising and lowering operators respectively.  }
\label{fig2}
\end{figure}

There are two triangle fermion graphs contributing which are crossed of
each other as shown in figure {\ref{fig2}}.
One may think that since `Up' and `Down' quarks carry opposite electric
charges the triangle graphs in which an `Up' quark couples to the photon and
the same graph in which the  `Down' plays that role give
opposite contributions to ${{\Delta k}_\gamma}(0),{{\Delta Q}_\gamma}(0)$.
If for the sake of the argument assume that all fermions are massless
,which is actually the case for the first two families,
this would  mean that the total fermionic contribution  is  proportional to
$$Trace \quad \{ Q  \} $$ which is well known to vanish
(anomaly cancellation condition).  This is stated in almost all previous
references and for this reason the contributions of the first two generations
of {\makebox{fermions}} are not considered.
Ignoring group factors the triangle graphs shown in figure \ref{fig2},
follow from each other under the interchanges
$$\alpha \rightleftharpoons \beta \quad,\quad
 {Q_{\mu}}\rightleftharpoons {Q_{\mu}},
{\Delta_{\mu}} \rightleftharpoons {-\Delta_{\mu}} $$
With the relavant group factors taken into account we get for the two graphs
\begin{eqnarray}
Graph (2a)\,&=&\,Tr(T_- T_+ Q)\,{V_{\mu\alpha\beta}}(Q,\Delta)\nonumber\\
Graph (2b)\,&=&\,Tr(T_+ T_- Q)\,{V_{\mu\beta\alpha}}(Q,-\Delta)\nonumber
\end{eqnarray}
where  the tensor  ${V_{\mu\alpha\beta}}(Q,\Delta)$ which also includes
the anomaly term is given by,
\begin{eqnarray}
{V_{\mu\alpha\beta}}(Q,\Delta)\,&=&\,{\alpha_{ 0}}
{\epsilon_{\alpha\beta\mu\lambda}}{\Delta^{\lambda}}    \nonumber  \\
&+&{\beta_1}g_{\alpha \beta}{\Delta_\mu}+
{\beta_2} (g_{\alpha \mu}{Q_\beta}-g_{\beta \mu}{Q_\alpha})+
{\beta_3} {\Delta_\mu} {Q_\alpha}{Q_\beta}+...   \nonumber
\end{eqnarray}
It is seen that anomaly term preserves its sign under the
interchange of indices and momenta given above unlike the rest of the terms
whose sign is flipped. This results to a total contribution,
$$\,Trace(\,Q\,\{T_-,T_+\})
{\epsilon^{\alpha\beta\mu\lambda}}{\Delta_{\lambda}}+
\,Trace(\,Q\,[T_-,T_+]\,)({\beta_1}g_{\alpha \beta}{\Delta_\mu}+...) $$
Thus the fermion contributions to the dipole/quadrupole moments are weighted
by
$$Trace \{ \, Q \, T^{(3)} \, \} $$
and the anomaly by
$$Trace \{ \, Q \, \}\,=0  $$
Thus `Up' and `Down' quarks yield same sign contributions despite the fact
that they carry opposite electric charges.This we think had been overlooked
in previous works.As a result the first two
generations , which we assume to have vanishing masses, yield nonzero
contributions to the dipole and quadrupole moments contrary to what has been
previoulsy claimed.
The fermionic contributions to ${{\Delta k}_\gamma},{{\Delta Q}_\gamma}$
of an $SU(2)$ doublet ${f \choose f^\prime}_L$
are thus given by,
\begin{eqnarray*}
&& \Delta k_{\gamma}=
  \frac {C_g}{2}   Q_{f^\prime}  \int_{0}^{1} dt
 {  t^4+(r_f-r_{f^\prime}-1)t^3+(2r_{f^\prime}-r_f)t^2   \over
     t^2+(r_{f^\prime}-r_f-1)t+r_f  }
   -   [f\rightleftharpoons {f^\prime}]        \\
&& \Delta Q_{\gamma} =
   \frac {2C_g}{3}  Q_{f^\prime} \int_{0}^{1} dt
  {   t^3(1-t)   \over
   t^2+(r_{f^\prime}-r_f-1)t+r_f     }
   -[f\rightleftharpoons {f^\prime}]           \\
&& (r_{f,{f^\prime}}  \equiv {(m_ { f,{f^\prime}} /M_W)}^2)
\end{eqnarray*}
The first two families yield,
$${{\Delta k}_\gamma}(0)=-1.334\quad,\quad{{\Delta Q}_\gamma}(0)
\simeq 1.776$$
always in units of $ {g^2}/{16{\pi^2}}$. Actually this is the largest
contributions of all sectors to ${{\Delta Q}_\gamma}(0)$   .
The third family contributes
$${{\Delta k}_\gamma}(0)\approx-.62\quad,\quad
{{\Delta Q}_\gamma}(0)=.145 $$
for $m_t = \, 170 GeV$ and $m_b\approx 5 GeV$ .
\subsubsection*{Higgs Bosons}
The Higgs sector of the MSSM is not like that of the $SM$. One needs
two Higgs doublets $H_1,H_2$ whose mass eigenstates
$A(neutral\, CP \, odd)$, $h_{ 0},H_{ 0},(neutrals\, CP \, even)$, and the
charged Higgses $H_{\pm}$ have the following masses,
$$
\begin{array}{ll}
A \,&:\, {m^2_A}= {m^2_1}+{m^2_2}   \\ \\
{H_0,h_0    }\, &:\,
{m_{ H,h }^2} =   \\
   &{\hspace{2mm}}  {\frac {1}{2}} \{  {(m^2_A+M_Z^2)}^2
\pm \sqrt {   {(m^2_A+M_Z^2)}^2 -4{M_Z^2} {m^2_A} {cos^2}(2\beta) }   \}  \\
  &  \\
{H_\pm}\, &: \, m^2_{H_\pm} =m^2_A+ M_W^2
\end{array}
$$
$h_{ 0},H_{ 0}$ are mixings of the fields $\xi_{1,2} \equiv Re{H_{1,2}^0}$
and this mixing is specified by the angle $\theta$
$$
\xi_2=(\cos \theta) h_0 +  (\sin \theta) H_0  , \quad
\xi_1=(\sin \theta) h_0 - (\cos \theta) H_0
$$
The field $h_0$  is predominantly   $\xi_1 $  if   $\sin \theta$
is close to unity and in that case it is the SM Higgs boson.
The contributions of a two Higgs model has been first discussed by
Couture et al. \cite{ng}; however no dependence  on the the mixing
angle  $\theta$  appears in their results.

At the tree level the neutral $h_0$
,the lightest of  ${H_0,h_0}$ , is lighter than the $Z$ gauge boson. However
radiative effects  due to the heavy top/stop system are substantial
resulting  to corrections that can push its mass  to values exceeding
$M_Z$ ( $m_h \approx 60 - 130 \, GeV $ for small $\tan \beta$).
This neutral yields the largest contributions of all Higgses involved.
 $H_{\pm},A,H_0$  have masses of the order of
$\cal{O}(M_{SUSY})$ , lying therefore in the $TeV$ range. Their contributions
to the dipole/quadrupole moments are much smaller.

The Higgs  contributions to the moments under discussion are  given by,
\begin{eqnarray*}
&&A\;:\qquad \Delta k_{\gamma}=D_2(R_A,R_+) \quad , \quad
            \Delta Q_{\gamma}=Q(R_A,R_+) \\
&&h_0\,:\qquad \Delta k_{\gamma}\:=\: \sin^2 \theta \: D_1\,(R_h)\:+\:
                                \cos^2 \theta \: D_2(R_h,R_+)  \\
&&\quad \qquad \hspace*{4.5mm} \Delta Q_{\gamma}\:=\: \sin^2 \theta \:
                             Q(R_h,1)\:+\: \cos^2 \theta \:Q(R_h,R_+)   \\
&&H_0\,:\quad \, \; \; As \,\, in\quad h_0\quad with \quad R_h \rightarrow R_H
\quad and
       \quad  \sin^2 \theta \rightleftharpoons  \cos^2 \theta   \\
&& \\
&&    R_a\equiv{(m_a/M_W)}^2 \quad a=h_0,H_0,A,H_{\pm}
\end{eqnarray*}
while the corresponding Standard Model Higgs contribution is,
\begin{eqnarray*}
&&\quad\quad   \Delta k_{\gamma}=D_1\,(\delta) \quad , \quad
               \Delta Q_{\gamma}=Q(\delta,1) \qquad
   (\,\,\delta\,=\,{(m_{Higgs}/M_W)}^2\,)
\end{eqnarray*}
In the equations above the functions $D_{1,2},Q$ are defined as,
\begin{eqnarray*}
&& \rhd  \quad \quad  D_1(r)\equiv{1\over 2} \int_{0}^{1} dt
        { 2t^4+(-2-r)t^3+(4+r)\,t^2   \over
        t^2+r(1-t)   }      \\
&&\qquad \quad   D_2(r,R)\equiv{1\over 2} \int_{0}^{1} dt
        { 2t^4+(-3-r+R)t^3+(1+r-R)t^2   \over
        t^2+(-1-r+R)t+r   }  \\
&&\qquad \quad   Q(r,R)\equiv{1\over 3} \int_{0}^{1} dt
        { t^3(1-t) \over
         t^2+(-1-r+R)t+r }
\end{eqnarray*}
Scanning the parameter space we found that the Higgs contributions to the
dipole and quadrupole moments receive values
$$
{\Delta \kappa_{\gamma}} \simeq 1. - .5 \quad ,\quad
{\Delta Q_{\gamma}} \simeq {\cal{O}}(10^{-2})
$$
\subsubsection*{Squarks-Sleptons}
This sector gives ${\cal{O}}(10^{-2})$ contributions to both
${\Delta \kappa_{\gamma}},{\Delta Q_{\gamma}}$ even in cases where
due to large mixings one of the stops turns out to be light. For a sfermion
$SU(2)$ doublet ${ {\tilde f} \choose {\tilde f}^\prime}_L$ these
contributions read as follows,
\begin{eqnarray*}
&& \Delta k_{\gamma}=
        - C_g  Q_{f^\prime} \sum_{i,j=1}^{2}
      {  ( {K_{i1}^{{\tilde f}} }  {K_{j1}^{{\tilde f}^\prime} }) }^2
      \int_{0}^{1}dt
   {   t^2(t-1)(2t-1+R_{{\tilde f}^\prime_j}- R_{{\tilde f}_i}) \over
      t^2+(R_{{\tilde f}^\prime_j}- R_{{\tilde f}_i}-1)t+
      R_{{\tilde f}_i}   }
      -[f\rightleftharpoons {f^\prime}]                    \\
&& \Delta Q_{\gamma}=
     - { { 2C_g} Q_{f^\prime}  \over 3   }  \sum_{i,j=1}^{2}
    {  ( {K_{i1}^{{\tilde f}} }  {K_{j1}^{{\tilde f}^\prime} }) }^2
     \int_{0}^{1}dt
   {  t^3(1-t)  \over
      t^2+(R_{{\tilde f}^\prime_j}- R_{{\tilde f}_i}-1)t+
      R_{{\tilde f}_i}   }
     -[f\rightleftharpoons {f^\prime}]                    \\
&&   \\
&&  \\
&&   R_{ {\tilde f}_i , {{\tilde  f}^\prime}_i }
     \equiv {(m_ { {\tilde f}_i , {{\tilde  f}^\prime}_i }  /M_W)}^2
 ;\quad m_ { {\tilde f}_i , {{\tilde  f}^\prime}_i } \quad are\,
 \,sfermion\,\,masses.
\end{eqnarray*}
The matrices ${\bf K}^ { {\tilde f},{\tilde f}^\prime }$ shown in the
expressions above diagonalize the sfermion mass matrices.
The calculation is complicated only by the presence of
${\tilde f}_L - {\tilde f}_R$ mixings due to the electroweak symmetry breaking
effects. In the absence of SUSY breaking  their
contributions to the quadrupole moment cancels against that of fermions as
they should.
\subsubsection*{Neutralinos-Charginos}
This is perhaps the most difficult sector to deal with due to
substantial mixings originating  from the EW symmetry  breaking effects.
In the chargino sector
the charged $SU(2)$ gauge fermions ${\tilde W}_{\pm}$ mix with the charged
Higgs fermions ${\tilde H}_{1}^{-},{\tilde H}_{2}^{+}$ through the mass matrix
$$
{\bf {{\cal M}_C}}=\left( \begin{array}{cc}
 M_2 & -g_2 \,v_2    \\
-g_2 \,v_1 & \mu
\end{array}\right)
$$
which is diagonalized by two unitary matrices $U,V$,
$$
U{{\cal M}_C}V^{\dag}=\, diag \{ m_1 \, , \,m_2 \}
$$
The mass eigenstates ${\tilde C}_1^+,{\tilde C}_2^+$ (Charginos) are
Dirac fermions  with masses  $m_1 \, , \,m_2$:
\begin{eqnarray}
     m_{1,2}^2&=&{1 \over 2} [M_2^2+{\mu}^2+2 M_W^2   \pm  \nonumber \\
 &   &    \sqrt{ ( (M_2-\mu)^2+2 M_W^2 (1+\sin 2\beta) )
      ( (M_2+\mu)^2+2 M_W^2 (1-\sin 2\beta) ) }]  \nonumber
\end{eqnarray}

In the  Neutralino sector,
the gauginos ${\tilde W}_3 \, , \, {\tilde B} $ and the neutral Higgsinos
${\tilde H}_{1}^{0},{\tilde H}_{2}^{0}$ get mixed with a mass matrix,
\begin{eqnarray}
{\bf {\cal M}_N}=\left( \begin{array}{cccc}
    M_1               &     0         &g^\prime v_1 / \sqrt 2
                                                 &  -g^\prime v_2/\sqrt 2 \\
    0                 &    M_2        &  -g v_1/\sqrt 2
                                                 &   g v_2/\sqrt 2       \\
g^\prime v_1/\sqrt 2  & -g v_1/\sqrt 2&       0
                                                 &        -\mu           \\
-g^\prime v_2/\sqrt 2  & g v_2/\sqrt 2 &      -\mu
                                                 &        0
\end{array}\right)  \nonumber
\end{eqnarray}
which is diagonalized by an orthogonal matrix $\cal O$ :
$$
{\cal O} {{\cal M}_N}{{\cal O}^T}=\, diagonal
$$
The eigenstates of the mass matrix ${\bf {\cal M}_N}$ are four Majorana
fermions ${{\tilde Z}_{\alpha}} \, , \, {\alpha}=1,2,3,4$.
%
Their Weak and Electromagnetic currents of these states are,
\begin{eqnarray}
{J^\mu_+}={ \sum_{\alpha,i}}{\bar {\tilde Z}}_\alpha \gamma^\mu
( R \, C^R_{\alpha i}+
L \, C^L_{\alpha i}) {\tilde C}_i^+  \quad,\quad
{J_{em}^\mu}={\sum_{i}} {\bar{\tilde C}}_i^+  \gamma^\mu   {\tilde C}_i^+
\nonumber
\end{eqnarray}
where the left and right handed couplings $C^{R,L}_{\alpha i}$  are given
in terms of $U,V,O$ matrices by,
\begin{eqnarray}
C^R_{\alpha i}=-{\frac {1}{\sqrt 2}} O_{3 \alpha} U^{*}_{i2}-
                                     O_{2 \alpha} U^{*}_{i1} \quad,\quad
C^L_{\alpha i}=+{\frac {1}{\sqrt 2}} O_{4 \alpha} V^{*}_{i2}-
                                     O_{2 \alpha} V^{*}_{i1} \nonumber
\end{eqnarray}
There is only one triangle graph contributing in this case
since the neutralini are Majoranna fermions yielding,
\begin{eqnarray*}
&& \Delta k_{\gamma}=- \sum_{i,\alpha} F_{\alpha i} \int_{0}^{1} dt
      {t^4+(R_{\alpha}-R_i-1)t^3+(2R_i-R_{\alpha})t^2  \over
      t^2+(R_i-R_{\alpha}-1)t+R_{\alpha}  }   \\
&&\qquad \quad +\sum_{i,\alpha}\,sign(m_im_\alpha)\,\,G_{\alpha i}\,
   \sqrt{R_{\alpha}R_i} \int_{0}^{1} dt {4t^2-2t \over
      t^2+(R_i-R_{\alpha}-1)t+R_{\alpha}  } \\
&& \Delta Q_{\gamma}=-{4 \over 3}\sum_{i,\alpha} F_{\alpha i} \int_{0}^{1} dt
        { t^3(1-t) \over t^2+(R_i-R_{\alpha}-1)t+R_{\alpha}  }  ,   \\
&& \quad \quad       (\, R_{\alpha ,i}\equiv{(m_{\alpha,i}/M_W)}^2\, ) \\   \\
\end{eqnarray*}
The prefactors in these formulae are :
$$
 F_{\alpha i}={|C^R_{\alpha i}|}^2+{|C^L_{\alpha i}|}^2\qquad ,\qquad
 G_{\alpha i}=( C^L_{\alpha i} \,{ {C^R_{\alpha i}}^*} +(h.c))
$$

The neutralino-chargino sector can accomodate light mass eigenstates and
in such a case the contributions to the dipole and quadrupole moments
are not in general suppressed. Actually we can
have a sizeable effect from this sector when there are
light neutralino-chargino states ($<M_W$) and this can  happen
provided the soft mass $M_{1/2}$ is smaller than $A_0,m_0$. In that case
$$ {{\Delta k}_\gamma}(0) ={\cal O}(1)\quad,\quad
{{\Delta Q}_\gamma}(0)={\cal O}(1) $$
When  $M_{1/2} \approx A_0,m_0 \sim {\cal O}(TeV)  $ both are
${\cal O}(10^{-2})$ or even {\makebox{smaller}} .
\section*{Numerical \quad analysis }
We  scanned the parameter space $A_0 , m_0 , M_{1/2}$ in the range
${\cal O}(100 GeV)$ to $ 1 TeV$. The space is divided into three regions:\\
\begin{itemize}
\item
$A_0 \simeq m_0 \simeq M_{1/2}$ \hspace{2mm} (comparable)
\item
$A_0 \simeq m_0 << M_{1/2}$ \hspace{2mm} ( Gaugino dominant )
\item
$A_0 \simeq m_0 >> M_{1/2}$ \hspace{2mm} ( Light gluinos )
\end{itemize}
For the top mass $m_t$ we considered values in the range ,
\begin{center}
       $ 130 GeV < m_t< 190 GeV $
\end{center}
We found that the MSSM predictions for the dipole and quadrupole moments
differ, in general, from those of the SM but they of the same order of
magnitude. Therefore supersymmetric structure can not be possibly probed
at LEP2. The order of magnitude of the contributions of the various sectors
, in units of $g^2/(16 \pi^2)$, are as shown in table \ref{tab3}.
\begin{table}
\caption{Order of magnitude contributions to $\Delta k_{\gamma}$,
$\Delta Q_{\gamma}$}
\label{tab3}
\begin{tabular}{ccc}
       &   $\Delta k_{\gamma}$    &  $\Delta Q_{\gamma}$         \\
\tableline
$ Gauge\quad bosons    $    & $+{\cal O}(1)$   &   $+{\cal O}(10^{-1}) $  \\
$ Matter\quad fermions $    & $-{\cal O}(1)$   &   $+{\cal O}(1) $        \\
$ Higgses     $            & $+{\cal O}(1)$ &
                                               $+{\cal O}(10^{-2})$ \\
$ {\tilde q} \, ,\, {\tilde l} $      &   $\pm{\cal O}(10^{-2})$  &
                                                 $\pm{\cal O}(10^{-2})$  \\
$ {\tilde Z }  \, , \,   {\tilde C}$ &  $\pm{\cal O}(10^{-2} - ?)$      &
                                           $\pm{\cal O}(10^{-2} - ?)$
\end{tabular}
\end{table}
Running the numerical routines we found regions of the  parameter space
allowing for light chargino and neutralino masses satisfying
$$m_{\tilde C}+m_{\tilde Z} \approx M_W$$
In such cases the values of
$\Delta k_{\gamma}$ , $\Delta Q_{\gamma}$ are substantially enhanced. In fact
in those cases the integrations over the Feynman parameter are of the form
$$\int_{0}^{1} f(t)dt/[(t-\alpha)^2+\epsilon^2]$$
with $0<\alpha<1$ and $\epsilon$ small (in situations like that we are
actually  close to an anomalous threshold). However even for such
relatively large contributions of this sector we can not have
values approaching the sensitivity limits of LEP2.
Only in a very limited region of the parameter space and when accidentally
the sum $m_{\tilde C}+m_{\tilde Z} $ turns out to be almost equal to
W - boson mass,
the chargino and neutralino contributions can be very large
saturating the sensitivity limits of LEP2. We disregard such large
contributions since they are not perturbatively trusted.
Even if it were not for that reason these cases are unnatural
occupying a very small portion of the available parameter space which is
further reduced if the lower experimental bound $m_{\tilde C} > 45 GeV$  on
the chargino mass is observed which
does not allow for arbitrarilly small  values of $M_{1/2}$). Therefore \\
{\em Although neutralinos and charginos may,
in some cases, yield
large  contributions approaching the sensitivity limits of LEP2 we do
not think that these cases are natural.}
\section*{Conclusions}
The main results of our analysis are :
\begin{itemize}
\item
The MSSM predictions for the Dipole and Quadrupole moments differ, in general,
from those of the SM but they are of the same order  of magnitute
$({ \cal{O}}(10^{-3}))$ in the entire parameter space $A_o , m_o , M_{1/2}$.
Experiments  should reach this level of accuracy for such differences to be
observed. Hence
deviations from the Standard Model predictions due to SUSY are unlikely to
be observed at LEP2.
\item
The Neutralino and Chargino sector is the principal source of
deviations from the SM predictions when this sector
involves light states $(<M_W)$.\\
 This occurs  when $M_{1/2}$  is light and for positive
values of $\mu > 0$ .
\item
The Sector of Neutralinos and Charginos may yield contributions to the Dipole
and Quadrupole moments whose magnitudes
saturates the sensitivity limits of LEP2. This happens
when $m_{\tilde C}\,+\,m_{\tilde Z}\,\simeq\,M_W$.
We consider these cases unnatural and perturbatively untrustworthy.
\item
To be of relevance for future collider experiments
the analysis should be extended to include values  $s \equiv 4Q^2 >4 M_W^2.$
The results of such an analysis will appear in a future publication
\cite{lahspa}.
\end{itemize}
\begin{table}
\caption{MSSM contributions to $\Delta k_{\gamma}$,$\Delta Q_{\gamma}$
for the inputs shown below. For comparison the SM predictions are shown
for $m_{Higgs}=50, 100/, and/, 300/, GeV$}
\label{tab4}
\begin{tabular}{ccccccc}
& &  $ m_t=160$ & &$\tan{\beta}=2$ & &   \\
& &  {$ A_0,m_0,M_{1/2}$}& =& 300, 300, 80 & &     \\
\tableline
 & & {$\Delta k_{\gamma}$ }& & & {$\Delta Q_{\gamma}$ }  & \\
   & $\mu > 0$  & & $\mu < 0$ &  $\mu > 0$ &  & $\mu < 0$  \\
\tableline
 $q,l$ & & -1.973 &   &   & 1.922  &    \\
$W,\gamma,Z$ & & 1.179          & &      &0.235  &   \\
$h_0,H_{\pm,0},A$ & & .946    & &          & .028 &  \\
${\tilde q},{\tilde l}$&.009& & -.035 &.027&  &.025 \\
${\tilde Z},{\tilde C}$&.697& &.026   &-.592& & -.170 \\
\tableline
Total &.859 & &.143   &   1.621&  & 2.041  \\
\tableline
SM  &  &  ${\Delta k_{\gamma}}^{SM}$ =& .188  & -.106,& -.449& \\
$(m_{Higgs}$ & 50,100,300 GeV) &
${\Delta Q_{\gamma}}^{SM}$= & 2.186 & 2.174 & 2.161&
\end{tabular}
\end{table}

\end{document}